# Deep Learning-Based Correction and Unmixing of Hyperspectral Images for Brain Tumor Surgery

*David Black[1*†], Jaidev Gill[2†], Andrew Xie[2†], Benoit Liquet[2,3,4], Antonio Di Ieva[4,5], Walter Stummer[6], Eric Suero Molina[4,5,6]**

1* Department of Electrical and Computer Engineering, University of British Columbia, Vancouver, BC, Canada.
2 Engineering Physics, University of British Columbia, Vancouver, BC, Canada.
[3]School of Mathematical and Physical Sciences, Macquarie University, Sydney, Australia
[4]Laboratoire de Mathématiques et de ses Applications, E2S-UPPA, Université de Pau & Pays de L'Adour, France
[4]Computational NeuroSurgery (CNS) Lab, Macquarie University, Sydney, New South Wales, Australia
5 Macquarie Medical School, Macquarie University, Sydney, NSW, Australia.
6 Department of Neurosurgery, University Hospital of Münster, Albert-Schweitzer-Campus 1, A1, D-48149, Münster, Germany.

*Corresponding authors. E-mails: dgblack@ece.ubc.ca;
eric.suero@ukmuenster.de;
†These authors contributed equally to this work

## Abstract

Hyperspectral Imaging (HSI) for fluorescence-guided brain tumor resection enables visualization of differences between tissues that are not distinguishable to humans. This augmentation can maximize brain tumor resection, improving patient outcomes. However, much of the processing in HSI uses simplified linear methods that are unable to capture the non-linear, wavelength-dependent phenomena that must be modeled for accurate recovery of fluorophore abundances. We therefore propose two deep learning models for correction and unmixing, which can account for the nonlinear effects and produce more accurate estimates of abundances. Both models use an autoencoder-like architecture to process the captured spectra. One is trained with protoporphyrin IX (PpIX) concentration labels. The other undergoes semi-supervised training, first learning hyperspectral unmixing self-supervised and then learning to correct fluorescence emission spectra for heterogeneous optical and geometric properties using a reference white-light reflectance spectrum in a few-shot manner.

The models were evaluated against phantom and pig brain data with known PpIX concentration; the supervised model achieved Pearson correlation coefficients (R values) between the known and computed PpIX concentrations of 0.997 and 0.990, respectively, whereas the classical approach achieved only 0.93 and 0.82. The semi-supervised approach's R values were 0.98 and 0.91, respectively. On human data, the semi-supervised model gives qualitatively more realistic results than the classical method, better removing bright spots of specular reflectance and reducing the variance in PpIX abundance over biopsies that should be relatively homogeneous. These results show promise for using deep learning to improve HSI in fluorescence-guided neurosurgery.

## Introduction

Identifying glioma margins is, due to their infiltrative growth, extremely difficult, if not impossible, during brain surgery. However, surgical adjuncts such as fluorescence guidance can maximize resection rates, thus improving patient outcomes[1,2]. 5-Aminolevulinic acid (5-ALA) is an FDA-approved tissue marker for high-grade glioma[3]. 5-ALA is administered orally four hours before induction of anesthesia for fluorescence-guided resection (FGR) of malignant gliomas; this drug is metabolized preferentially in tumor cells to protoporphyrin IX (PpIX), a precursor on the heme synthesis pathway[4]. PpIX fluoresces bright red, with a primary peak at 634 nm, when excited with blue light at 405 nm. In this way, tumors that are otherwise difficult to distinguish from healthy tissue can sometimes be identified by their red glow under blue illumination. This allows for a more complete resection and thus improved progression and overall survival[2,5]. However, the fluorescence is often not visible in lower-grade glioma or in infiltrating margins of tumors[6,7]. In these cases, the PpIX fluoresces at a similar intensity to other endogenous fluorophores, known as autofluorescence.

Hyperspectral imaging (HSI) is, therefore, an active research area, as it allows the PpIX content to be isolated from autofluorescence. HSI systems capture three-dimensional data cubes in which each slice is an image of the scene at a particular wavelength. A fluorescence intensity spectrum is obtained by tracing a pixel through the cube's third dimension. Thus, the emission spectrum of light is measured at every pixel[8]. Within each pixel is a specific combination of fluorescing molecules or fluorophores. Assuming a linear model that neglects multiple scattering[9] and other effects, the measured fluorescence spectrum ($\Phi_{Fluo}$) is thus a linear combination of the emission spectra of K potential present fluorophores ($\Phi_{Spec,k}$), also called endmember spectra[10], as shown in Equation 1 (ignoring noise). With a priori knowledge of the endmember spectra, linear regression techniques have been employed to determine the relative abundances ($c_k$) of the endmembers[11].

$$\Phi_{Fluo} = \sum_{k=1}^{K} c_k \, \Phi_{Spec,k} \qquad 1$$

Recent advances in hyperspectral imaging (HSI) for fluorescence-guided surgery have increased our ability to detect tumor regions[8,12] and even classify tissue types based on the endmember abundances[13,14]. They have also been used to study 5-ALA dosage[7] and timing[6], and to improve the imaging devices[15–17]. However, these computations are extremely sensitive to the autofluorescence, as well as to artifacts from the optical and topographic properties of the tissue and camera system. To mitigate the latter issue, the measured fluorescence spectra are corrected to account for heterogeneous absorption and scattering by comparing them to the measured spectra under white light illumination at the same location. Correction by dual-band normalization involves integrating over two portions of these spectra, raising one to an empirical exponent, and multiplying them to determine a scaling factor[18]. While effective in phantoms[19], we have found this method to be of limited use in patient data[12]. The pixels are also corrected for their distance from the objective lens, with further pixels appearing dimmer than closer ones[20,21]. Other methods are relatively simplistic, linear, and not based on human data[22]. They are thus unable to account for nonlinear effects such as multiple scattering[9], the dual photostates of PpIX[4,23], and fluorescence variation due to pH and tumor microenvironment[12], nor can they fully correct for the inhomogeneous optical properties of the tissue[12]. These effects may also include wavelength-dependent absorption and scattering variations, which are unmodeled when using a single scaling factor. An example of attenuation correction is shown in Fig. 1.

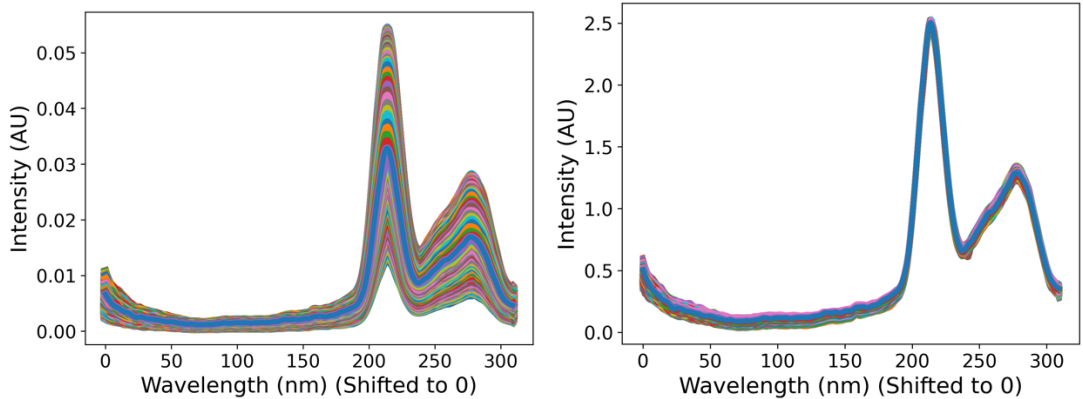

**Figure 1.** Typical attenuation correction of measured spectra from a phantom of constant PpIX concentration. On the left are the raw spectra, with large variance, and on the right are the normalized ones after correction. The variance in the magnitudes is greatly decreased.

Once the spectra are corrected for optical and topological variations, they must be unmixed into the endmember abundances. In 5-ALA-mediated fluorescence-guided tumor surgery, these likely include the two photostates of PpIX[4,23], called $PpIX_{620}$ and $PpIX_{634}$, as well as autofluorescence from flavins, lipofuscin, NADH, melanin, collagen, and elastin[11,24], though there are usually only 3 or 4 endmembers present in any given spectrum[25]. Previous work has commonly used non-negative least squares (NNLS) regression[4,11–13,26,27]. This is simple and fast and guarantees non-negative abundances. Three example unmixings using NNLS are shown in Fig. 2. Others have proposed Poisson regression[28] to account for the theoretically Poisson-distributed photon emissions[29], or various sparse methods to reduce overfitting and enforce the fact that there are usually only few fluorophores present in each pixel[10]. However, as mentioned before, all of these methods assume linearity in the combination of the endmember spectra. Furthermore, they rely on the normalization to be accurate.

Thus, it would be beneficial to perform the normalization and unmixing in a single step process that can handle the nonlinearity and complexity of the physical, optical, and biological systems described above. For this purpose, deep learning is particularly well suited, mainly as each HSI measurement produces a large volume of high-dimensional data. Indeed, deep learning has been explored in detail for HSI, as reviewed by Jia et al.[30], and for medical applications specifically[31,32]. For brain tumor resection, the technique is very promising[33] and several studies have used support vector machines (SVMs), random forest models, and simple convolutional neural networks (CNNs) to perform segmentation and classification of tissues in vivo[34–36]. Other approaches include majority voting-based fusions of k-nearest neighbors (KNN), hierarchical k-means clustering, and dimensionality reduction techniques such as principal component analysis (PCA) or t-distributed stochastic neighbor embedding (t-SNE)[37,38]. These papers used 61 images from 34 patients with a resulting median macro F1-Score of 70% in detecting tumors. Rinesh et al. used KNNs and multilayer perceptrons (MLPs)[39]. The HELICoiD (Hyperspectral Imaging Cancer Detection) dataset[40], which consists of 36 data cubes from 22 patients, has been widely used. For instance, Manni et al. achieved 80% accuracy in classifying tumor, healthy tissue, and blood vessels using a CNN[41], and Hao et al. combined different deep learning architectures in a multi-step pipeline to reach 96% accuracy in glioblastoma identification[42]. Other methods used pathological slides[43], with most mentioned based on small datasets[44].

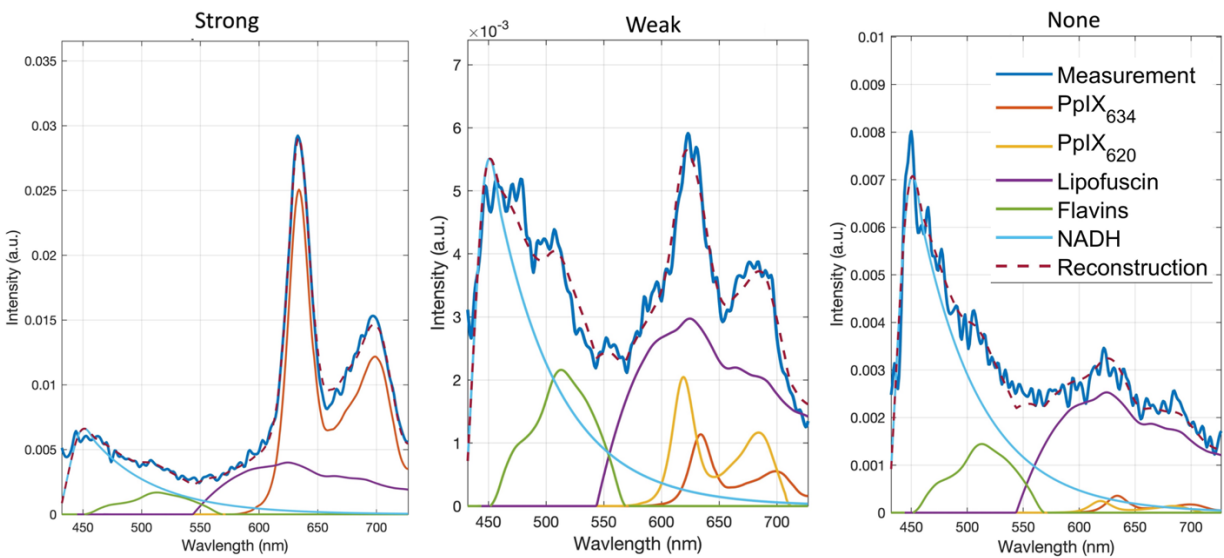

**Figure 2.** Sample unmixing of three spectra with strong, weak, and effectively no PpIX content. The blue solid line is the measured spectrum, while the purple dashed line is the fit. The other spectra are the endmembers, scaled according to their abundance and summed to create the fitted spectrum. Unmixing enables recovery of PpIX abundance despite autofluorescence.

Given the small datasets, many of these papers did not have sufficiently good results to be clinically useful yet and likely do not generalize very well. This is partly due to the cost of labeling many hyperspectral images. As a result, modern architectures for medical image segmentation, such as U-Net[45], V-Net[46], or graph neural networks (GNNs)[47], have seen little use. Autoencoders[48] or generative adversarial networks (GANs)[49] can use unsupervised learning for certain tasks to avoid the labeling problem but require large volumes of data. Jia et al. describe some approaches to overcome the lack of data in HSI[50]. In addition, these papers all represent end-to-end attempts to take a raw data cube containing high-grade glioma and output a segmentation. This approach is unlikely to generalize well to other devices, hospitals, or tumor types. Instead, a more fine-grained method may generalize better, in which the core steps of the process are individually optimized and rooted in the physics of the system. These steps include image acquisition, normalization, unmixing, and interpretation of endmember abundances. The surrounding elements of device-specific processing can be kept separate. This separation also enables more flexible use of the results. For example, the endmember abundances may be used to identify tumor tissue, but also to classify the tumor type or provide information about biomarkers such as isocitrate dehydrogenase (IDH) mutation, which is clinically highly relevant[13].

As described before, classical methods for unmixing have some limitations. Therefore, research has explored deep learning-based unmixing. Zhang et al. successfully applied CNNs to this task to obtain endmember abundances on four open-source agricultural HSI datasets[51]. No similarly large dataset is available for brain surgery. Wang et al. used CNNs to obtain slightly better performance than nonnegative matrix factorization on simulated and real geological HSI data[52]. Others have used fully connected MLPs[53], CNNs[54], and auto-associative neural networks[55] to unmix spectra without prior knowledge of the endmembers. However, these are not as effective when the endmember spectra are known, as in our case. An attractive solution called the endmember-guided unmixing network (EGU-Net) used autoencoders in a Siamese configuration to enforce certain relevant constraints, with good results[56]. A review on deep learning-based unmixing by Bhatt and Joshi shows that existing work is relatively minimal and preliminary[57]. Much of the research does not use a priori known endmember spectra, and to the authors' knowledge, none focuses on the attenuation correction, on neurosurgery, or on fluorescence-assisted HSI.

This paper, therefore, describes a novel method of deep learning-based normalization and unmixing of HSI data cubes for fluorescence-guided resection of brain tumors. This improves on classical methods, can fit into any HSI pipeline in brain surgery, and gives generalizability and flexibility in the use of the endmember abundances. This is facilitated by the first use, to the authors' knowledge, of modern architectures, including deep auto-encoders and residual networks[58] in HSI for brain tumor surgery. It is also the first use of a large and broadly diverse dataset for deep learning in HSI for neurosurgery, including 184 patients and 891 fluorescence HSI data cubes from 12 tumor types, all four World Health Organization (WHO) grades, with IDH mutant and wildtype samples, and labeled solid tumor, infiltrating zone, and reactive brain ("healthy") tissue. The models are optimized using phantoms and pig brain homogenate data with known PpIX concentration. Still, due to the physical underpinnings of the design, we are able to show not only better quantitative results on these distributions but also improvements in generalizing to human data as well.

## Methods
### Architecture
Two neural network architectures were developed and tested: a supervised model we called ACU-Net and a novel unsupervised or semi-supervised autoencoder model called ACU-SA, inspired by EGU-Net[56]. As described in the Introduction, CNNs are popular for hyperspectral image analysis. Given the data's characteristics, we employ a 1D deep Convolutional Neural Network (1D CNN) architecture. This choice is particularly suited since neighboring wavelengths exhibit more similar distortions than those farther apart. Furthermore, for the fluorescence spectrum, correlations are more pronounced with adjacent wavelengths in the white light reflectance spectrum than with distant ones. This spatial and spectral correlation aligns well with the inductive bias inherent in CNNs, making them an apt choice for our model. Furthermore, we leverage residual connections, which allow for bypasses of certain layers[59] and have been demonstrated to be a robust heuristic choice that improves the quality of learned features[60].

For both models, the input data consists of $X \in \mathbb{R}^{n \times m \times 2}$ where $n$ is the number of spectra and $m$ is the number of wavelength samples in each spectrum. We use the fluorescence emission spectrum $\Phi_{Fluo} \in \mathbb{R}^m$, which is captured after exciting the region with light at $\lambda = 405$ nm, and the white light reflectance spectrum $\Phi_{Ref} \in \mathbb{R}^m$, which is captured after exciting the region with broadband white light. The two spectra are stacked to form a two-channel input spectrum, which utilizes the locality bias of the CNN. Let $K$ be the number of known endmember spectra. The matrix whose columns are the endmember spectra is $B = [\Phi_{spec,1} \quad \cdots \quad \Phi_{spec,k}] \in \mathbb{R}^{m \times K}$.

The HSI data attenuation correction aims to correct the fluorescence emission spectra so that those originating from samples with equal fluorophore concentration have equal magnitudes irrespective of local optical or geometric properties. In other words, the goal becomes to minimize the variation between spectra. Suppose there is an ideal corrected spectrum, $\Phi_c$, which is the pure emission of the fluorophores with all effects corrected for. Then the correction seeks to minimize the variance between the predicted fluorescence spectra $\widehat{\Phi}_{fluo,i}$, and $\Phi_c$, $i \in [1, n]$. Thus, we use the mean squared error (MSE = $\frac{1}{n}\sum_{i=1}^{n}\|\widehat{\Phi}_{fluo,i} - \Phi_c\|^2$) for the proposed models when predicting the true fluorescence emission spectra. Note $MSE(x) = Bias(x)^2 + Var(x)$ [61], so minimising this objective function does indeed minimize the variance between the predicted and the true normalized spectra. For models in which the abundances are output rather than reconstructed spectra, i.e. the output is the $c_k$ from Equation 1 rather than the $\Phi_{fluo}$, MSE is also used.

*ACU-Net*
The Attenuation Correction and Unmixing Network (ACU-Net) is a 1D CNN with four residual blocks, each containing 2-3 same-convolutions, each followed by a small max pooling layer to reduce the dimensions of the feature maps. Between each residual block, there is a convolution layer that approximately doubles the number of feature channels. A kernel size of 5 is used in the early layers, and 3 in the later ones. The output of the convolutional layers is inputted to three fully connected layers. The architecture is shown in Fig. 3. The white light and fluorescence emission spectra are stacked, so convolutions are performed together, as described previously.

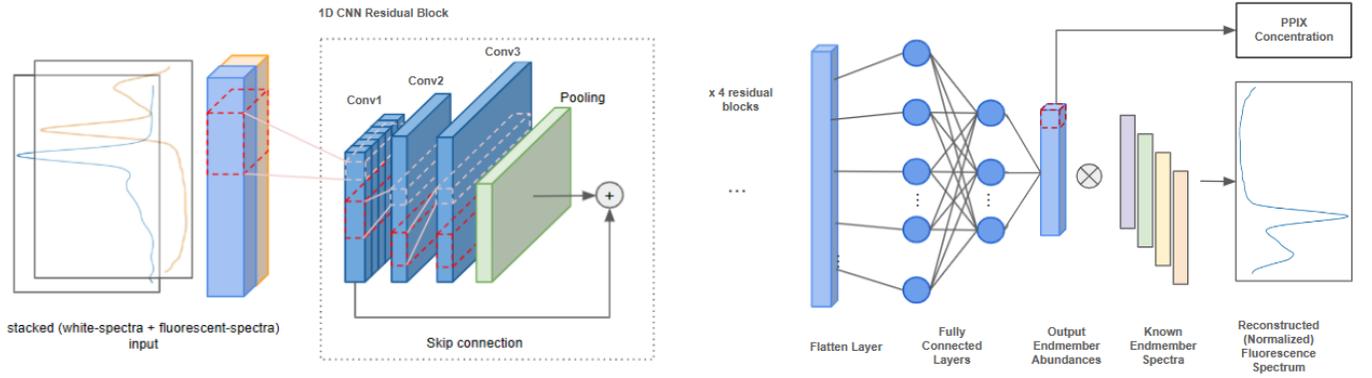

**Figure 3.** Spectrally-informed attenuation correction and hyperspectral unmixing network (ACU-Net) architecture.

The goal of the ACU-Net is to learn the mapping $f: \mathbb{R}^{m \times n \times 2} \to \mathbb{R}^K$ from the raw measured spectrum to the absolute endmember abundance vector, $z$, which includes $PpIX_{620}$, $PpIX_{634}$, and three primary autofluorescence sources: lipofuscin, NADH, and flavins[11]. Other autofluorescence may be present[24], but these 5 spectra have been shown to fit well[11]. This mapping is shown in Equation 2. The ground truth absolute PpIX concentration, denoted $c_{PPIX}$, is known for the phantoms and is known on average for the pig brain homogenate, as described by Walke et al.[12].

$$z = f(\Phi_{Fluo}, \Phi_{Ref}), \quad z \in \mathbb{R}^K \qquad 2$$

Ground truth abundances are not, however, known for human data. Thus, a second loss - the reconstruction loss - is also considered with the aim of better generalization to human data. Let the relative abundance vector be $\hat{z} = \frac{z}{\|z\|_2}$. With these, we can define the normalized reconstructed spectrum $\widehat{\Phi}_{Fluo} = \sum_{k=1}^{K} \hat{z}_k \Phi_{spec,k} = B\hat{z}$, which should be as close as possible to the true corrected spectrum, $\Phi_c$ described above. The mapping $f$ thus also aims to minimize $MSE(\widehat{\Phi}_{fluo,i} - \Phi_c)$. There is no known ground truth spectrum $\Phi_c$. Instead, ACU-Net uses the non-corrected $\Phi_{Fluo}$, hypothesizing that by training on a large a diverse dataset, $\widehat{\Phi}_{fluo}$ will converge towards an average representation that best characterizes $\Phi_c$. We observe in the Results section that this indeed occurs. Additionally, it is essential to note that the learned $\widehat{\Phi}_{fluo,i}$ will not fit as precisely as methods employing least squares (LS), which are mathematically optimal. However, by utilizing a deep neural network (DNN), we aim to more effectively learn the corrected fluorescence spectrum $\Phi_C$, and the abundances underlying the noisy measurement.

Using a rectified linear unit (ReLU) activation function at the output of the final layer enforces the non-negativity constraint on the relative abundance values. Finally, we use a weighted loss to train the model to minimize both the error in predicted concentration and the reconstruction error. Since the two objectives are of different scales and it is unknown how the structure of our architecture may affect the learning, the loss weights are also parameterized by considering the homoscedastic uncertainty of each task as outlined by Kendal et al.[62]. This formulation of the loss function for deep multi-task learning has been shown to be effective for multiple regression and classification objectives for CNNs in many computer vision tasks. Denoting $\sigma_C$ and $\sigma_{rec}$ as the learned parameters for weighing the concentration prediction and spectrum reconstruction components of the architecture, we write the total loss function for one measured spectrum in Equation 3.

$$L = \frac{1}{2\sigma_C^2}(z_1 - c_{PPIX})^2 + \frac{1}{2\sigma_{rec}^2}\|Bz - \Phi_{fluo}\|_2^2 + \log(\sigma_C \sigma_{rec})$$
$$z = f(\Phi_{Fluo}, \Phi_{Ref})$$
3

*ACU-SA*

The challenge with ACU-Net is that it requires ground-truth abundance labels, which are only available for phantom data. Therefore, we also propose a semi-supervised model. Attenuation Correction and Unmixing by a Spectrally-informed Autoencoder (ACU-SA) is similar to the EGU-Net[56], using an endmember-guided unsupervised approach to the unmixing process. ACU-SA consists of two main components: one for hyperspectral unmixing (HU) and one explicitly for normalization. The HU portion consists of a Siamese autoencoder architecture, as shown in Fig. 4, outlined in green. The objective of this portion is to learn a mapping $f: \mathbb{R}^{n \times m \times 1} \to \mathbb{R}^K$, from the normalized fluorescence spectrum to the absolute endmember abundances, like ACU-Net. However, unlike ACU-Net, this portion takes the attenuation-corrected fluorescence emission spectrum as an input rather than the stacked raw spectra and, through its autoencoder structure, unmixes and reconstructs it. The HU component has the same architecture as the ACU-Net. Then, ACU-SA also includes a standalone CNN normalization model (blue outline in Fig. 4) whose objective is to learn the mapping $g: \mathbb{R}^{n \times m \times 2} \to \mathbb{R}^{n \times m \times 1}$, from the two captured spectra to an intermediate representation which we train to be the normalized/corrected fluorescence spectrum. Together, the normalization model takes the stacked white and fluorescence spectra, performs the attenuation correction, and feeds into the HU autoencoder network, which unmixes it into the absolute endmember abundances. The normalization model we used is a shallow 1DCNN with four convolutional layers with no residual blocks.

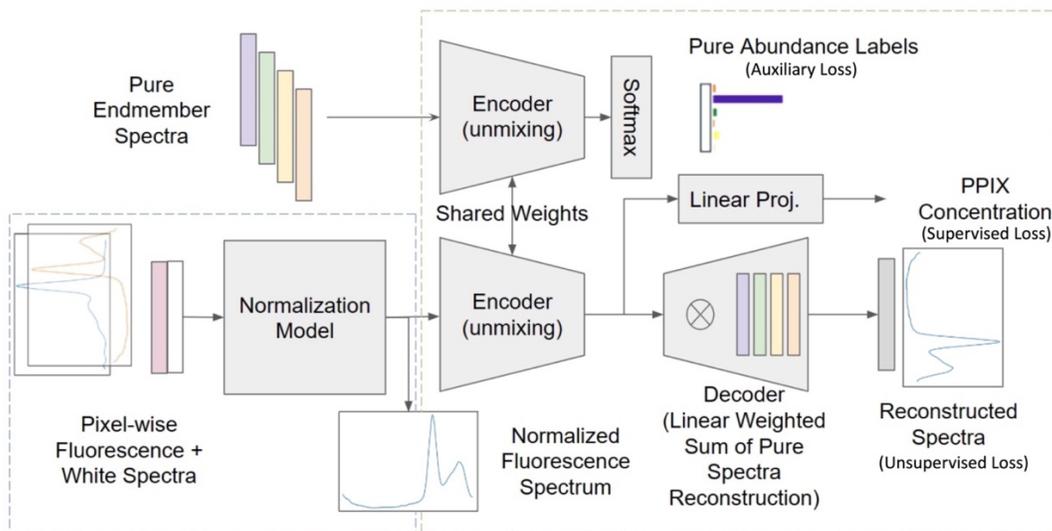

**Figure 4.** Endmember-guided normalization (blue outline) and unmixing (green outline) network for unsupervised or semi-supervised learning through an autoencoder architecture

For supervised learning, the output embeddings from the encoder can be compared to known abundances using the MSE. Otherwise, the decoder reconstructs the spectrum from the abundance values so it can be compared to the input spectrum to obtain an unsupervised reconstruction loss. As with ACU-Net, the decoder uses the output embeddings as weights in the linear combination from Equation 1. Thus, the decoder has fixed parameter weights to ensure the encoders embeddings represent the real endmember abundances.

A twin encoder with shared weights to the HU encoder is used with a SoftMax output and evaluated with a cross entropy loss. The pure endmember spectra are input to this network, and the output should ideally be a one-hot vector. For example, if the second endmember is input, the unmixing should output zero for all the endmembers except the second, which should be one. In this way, the independence of the endmember spectra are enforced, and we ensure that the output embeddings each correspond to only one endmember. This conditions the network on our a priori knowledge of the endmember spectra and has been shown to be effective in deep neural networks for HU[56].

ACU-SA is trained in two stages. First, the HU network is trained to learn $f$ for the pig brain and homogenate datasets. Since this stage is fully self-supervised, we can augment the training data with synthetic data composed by creating random linear combinations of the known endmember spectra to help the HU module learn unmixing more effectively, and we can use unlabeled human data. The loss function used for training the HU is given in Equation 4, where $\hat{e}_k \in \mathbb{R}^K$ is all zeros with a 1 in the $k^{th}$ element.

$$L_{HU} = \frac{1}{2K\sigma_{EG}^2} \sum_{k=1}^{K} CE\left(f(\Phi_{spec,i})\right) + \frac{1}{2\sigma_{rec}^2}\left\|Bz - \hat{\Phi}_{fluo}\right\|^2 + \log(\sigma_{EG}\sigma_{rec}) \quad 4$$

$$z = f(\hat{\Phi}_{fluo})$$

$$CE(\phi) = \log \frac{e^{\phi^T \hat{e}_k}}{\sum_{j \in \{1,\dots,K\}} e^{\phi^T \hat{e}_j}}$$

Here $\sigma_{EG}$ and $\sigma_{rec}$ are again learned loss weightings as used in ACU-Net. For the second stage of training, the weights of the HU module are frozen, and the normalization module is attached. Then, given a much smaller amount of data labeled with their PpIX concentrations, the full network can be trained, optimizing only the weights for the normalization module. The loss function for this stage is shown in Equation 5, where $[x]_i$ represents the $i^{th}$ element of vector $x$.

$$L = \left(\left[f\left(g(\Phi_{fluo},\Phi_{ref})\right)\right]_1 - c_{PpIX}\right)^2 \quad 5$$

**Dataset**

Since the presented models require a mix of labeled and unlabeled data, three datasets were used in this paper: (1) brain tissue phantoms were created using known concentrations of PpIX, (2) homogenized pig brain was spiked with known concentrations of PpIX, and (3) human brain tumor tissue was extracted during surgery and imaged ex vivo. All samples were measured on the same HSI device at the University Hospital of Münster, described below. Of the human data, 1000 example spectra were randomly sampled and are plotted in Fig. 5.

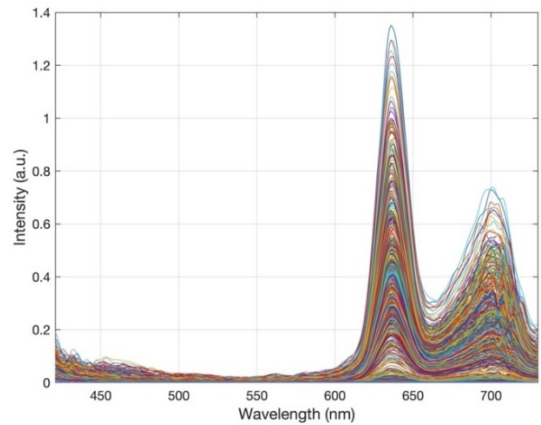

**Figure 5.** 1000 typical human fluorescence spectra were randomly sampled from the dataset of 555666 total spectra. These show clear PpIX content and vary widely in magnitude.

For phantoms, PpIX was mixed with Intralipid 20% (Fresenius Kabi GmbH, Bad Homburg, Germany) and red dye (McCormick, Baltimore, USA) in dimethyl sulfoxide (DMSO; Merck KGaA, Darmstadt, Germany) solvent to simulate the scattering and absorption, respectively, in human tissue, as described by Valdes et al[18,19]. The PpIX concentrations were (0.0, 0.2, 0.6, 1.25, 2.5 µg/ml). By varying the other ingredients, the following optical properties were achieved: absorption at 405 nm: $\mu_{a,\ 405\ nm}$ = 18, 42, 60 cm$^{-1}$; reduced scattering at 635 nm: $\mu'_{s,\ 635\ nm}$ = 8.7, 11.6, 14.5 cm$^{-1}$.

For pig brain homogenate (PBH), pig brain was obtained from a local butcher and separated into anatomical sections of cerebrum, cerebellum, hypothalamus, and brain stem/spinal cord. The tissue was washed with distilled water, cut into 10 × 10 × 10 mm pieces, and homogenized using a blender (VDI 12, VWR International, Hannover, Germany). The pH was controlled using 0.5 M tris(hydroxymethyl)aminomethane (Tris-base, Serva, Heidelberg, Germany) buffer and hydrochloric acid (HCl, Honeywell Riedel–de Haen, Seelze, Germany). For each sample, 200 to 600 mg of the homogenates were spiked with PpIX (Enzo Life Sciences GmbH, Lörrach, Germany) stock solution (300 pmol/µl in DMSO) to the desired concentrations (0.0, 0.5, 0.75, 1.0, 2.0, 3.0 and 4.0 pmol/mg) and homogenized using a vortex mixer. The PBH samples were placed in a Petri dish, making samples of about 4 × 4 × 2 mm. Approval for experiments with pig brains was given by the Health and Veterinary Office Münster (Reg.-No. 05 515 1052 21).

Finally, the human data was measured over 6 years (2018-2023) at the University Hospital of Münster, Münster, Germany. Patients undergoing surgery for various brain tumors were given a standard dose of 20 mg/kg of 5-ALA (Gliolan®, medac, Wedel, Germany) orally four hours before induction of anesthesia. Tissue resected by the surgeons was immediately taken to the HSI device and imaged ex vivo before being given to pathology.
**Ethics:** Informed consent was obtained from each individual in this patient collective. All procedures performed in studies were in accordance with the ethical standards of the institutional and/or national research committee and with the 1964 Helsinki declaration and its later amendments or comparable ethical standards. Data collection and scientific use of biopsies had previously been approved by the ethics committee of the University of Münster.

All the samples from these sources were imaged using an HSI device previously described[6–8,11–13]. The sample was illuminated with white light from a xenon light source to capture the white light spectra, blue light from a 405 nm LED for the fluorescence spectra, and not at all for dark spectra, which were used to remove the dark noise of the camera sensor. The reflected and emitted light was captured with a ZEISS Opmi Pico microscope (Carl Zeiss Meditec AG, Oberkochen, Germany) and passed through several low and high-pass filters to remove, for example, the brightly reflected blue excitation light. The light then passed through a liquid crystal tunable filter (Meadowlark Optics, Longmont, CO, USA) to a scientific metal oxide semiconductor (sCMOS) camera (PCO.Edge, Excelitas Technologies, Waltham, MA, USA). Data cubes were captured by sweeping the filter through the visible range and capturing a 2048 x 2048 pixel grayscale image at every sampling wavelength. Each image had a 500 ms exposure time to ensure good signal-to-noise ratio even from faint fluorescence.

Once captured, each data cube contained the sample of interest surrounded by background of the slide. Extracting the spectra from only the sample by manual segmentation is tedious, so classical computer vision techniques of edge and blob detection and morphological opening were used to detect the sample automatically. This was later augmented using a Detectron 2 model trained on our images[63]. Within these selected areas, regions of 10 x 10 pixels were averaged to increase the signal-to-noise ratio, and as many non-overlapping regions as possible were extracted from the biopsy to ensure independent data samples. The spectra were then corrected for the filter transmission curves and wavelength-dependent sensitivity of the camera. Approximately 500 to 1000 spectra were measured from each biopsy.

In total, data cubes were measured for 891 biopsies from 184 patients, resulting in 555666 human brain tumor spectra. The phantom data consisted of 9277 spectra, and the PBH samples were large and constituted 198816 spectra. The full breakdown of the data is shown in Table 1.

| Category | # of Data Cubes | Category | # of Data Cubes |
|---|---|---|---|
| *Tissue Type* | 632 | *Margins (Gliomas)* | 288 |
| Pilocytic Astrocytoma | 5 | Reactively altered brain tissue | 100 |
| Diffuse Astrocytoma | 57 | Infiltrating zone | 57 |
| Anaplastic Astrocytoma | 51 | Solid tumor | 131 |
| Glioblastoma | 410 | | |
| Grade II Oligodendroglioma | 24 | *WHO Grade (Gliomas)* | 571 |
| Ganglioglioma | 4 | Grade I | 9 |
| Medulloblastoma | 6 | Grade II | 84 |
| Anaplastic Ependymoma | 8 | Grade III | 57 |
| Anaplastic Oligodendroglioma | 4 | Grade IV | 421 |
| Meningioma | 37 | | |
| Metastasis | 6 | ***IDH Classification*** | *411* |
| Radiation Necrosis | 20 | Mutant | 126 |
| | | Wildtype | 285 |
| ***Phantoms*** | 5 (9277 spectra) | | |
| | | ***Pig Brain Homogenate*** | 12 (198816 spectra) |

***Table 1.*** *Breakdown of analyzed data. 891 hyperspectral data cubes of ex vivo tissue from fluorescence-guided surgeries of 184 patients and phantom and pig brain homogenate measurements were evaluated. Various categories are listed within the human data to show the diversity of the dataset.*

**Experiments**

Various tests were performed to determine the performance of the models on the dataset. As a baseline method or benchmark, the classical attenuation correction and unmixing procedures described in the Introduction were used, including dual-band normalization and nonnegative least squares unmixing. To evaluate the performance, we used not only the MSE of the reconstructed spectra or calculated abundance vectors, but primarily the correlation coefficient (R) between the measured and ground truth PpIX concentrations. This should ideally be linear, so an R as close to 1 as possible is desired. In this way, the method can be calibrated with a single scaling factor.

To test the fully supervised ACU-Net, it was necessary to use the phantoms and PBH data, which had ground truth labels. It was possible to train the ACU-SA on human data. However, assessing its performance was difficult without known abundances, and thus comparing methods was impossible. Therefore, the ACU-SA was also trained on the phantom and PBH data for quantitative evaluation before using the human data for a qualitative analysis. For training the ACU-SA, each dataset was split into 85% and 15% into training and testing sets, respectively. All results presented below are on test data unseen during training. The models were trained using the AdamW optimizer with an adaptive learning rate that decreases on training loss plateau. No hyperparameter tuning was done.

## **Results**

Fig. 6 shows the true and predicted PpIX concentration in PBH data using ACU-Net in contrast to the former approach. The data has lower variance, indicating that the attenuation correction is effective. Additionally, the unmixed PpIX abundances are linear with the known abundances. Thus, the unmixing is also effective. In fact, the coefficient of determination for the PBH data was 0.97 using ACU-Net, compared to 0.82 with the benchmark method. The R value was similarly strongly improved for phantom data, as shown in Table 3.

This shows that the supervised deep learning method can outperform classical methods. However, the unsupervised ACU-SA method, too, shows a marked improvement in performance compared to the benchmark, with R values comparable to the supervised model. All the results for phantom and PBH data with the three methods are shown in Table 3.

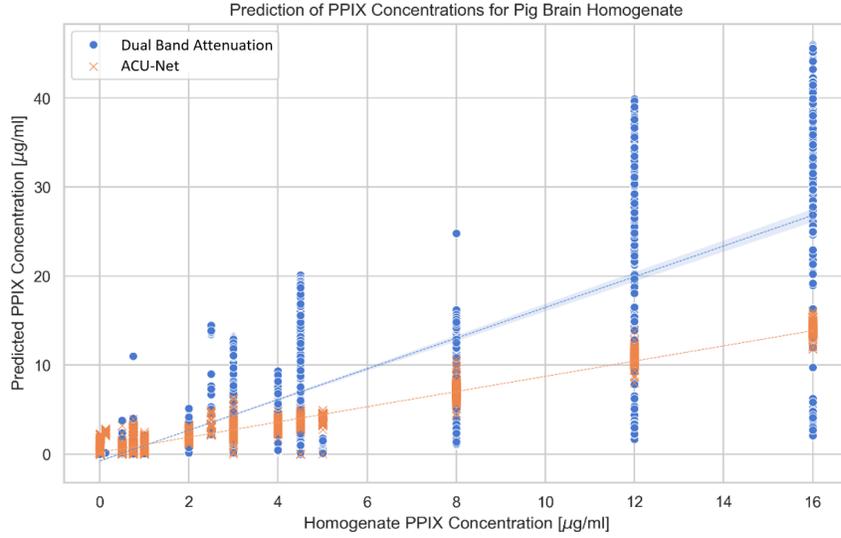

**Figure 6.** Linearity and variance of ACU-Net normalization and unmixing compared to the classical method. We see greatly improved linearity at a much lower variance and, consequently, a higher coefficient of determination.

*Table 3.* Comparison of proposed end-to-end learning-based normalization and unmixing compared to the benchmark dual band normalization followed by nonnegative least squares unmixing. The coefficient of determination between known and computed PpIX concentration is used for consistency with previous normalization work[61,71].

|  | *Benchmark* | *ACU-Net* | *ACU-SA* |
|---|---|---|---|
| *Phantom Data* | R = 0.93 | R = **0.997** | R = 0.98 |
| *Pig Brain Homogenate* | R = 0.82 | R = **0.99** | R = 0.91 |

Though the phantom and PBH results are promising, the critical question is whether the same results hold true in human data. While we currently do not have the true endmember abundances and thus cannot assess the performance quantitatively, we can observe that the average MSE reconstruction error is $2.72 \times 10^{-4}$ AU, comparable to that of the benchmark method ($9.63 \times 10^{-5}$ AU). Note that the NNLS unmixing minimizes the sum of squared errors, so it is not possible to outperform it in this metric. It shows, however, that the model outputs are reasonable and close to optimal.

Furthermore, differences in the output PpIX concentration maps are observed. In many cases where strong spots of specular reflection caused anomalous results in the dual-band normalization[12], the ACU-SA can remove the artifacts. This may be because the white light spectra in these cases were sometimes saturated, so the dual-band normalization would not sufficiently compensate, while a deep learning approach can better cope. In addition, previous papers have noted the difficulty of calibrating the unmixing output due to the nonlinear nature of PpIX fluorescence and the presence of more than one fluorescing state with different peak wavelengths[4,12,23]. These factors lead, with the previous method, to unexpectedly large output PpIX concentrations in many cases. However, with the ACU-SA, the values appear far more reasonable, adhering more to expected values with less extreme variation. These factors are illustrated in Fig. 7 and suggest that the deep learning approaches may have several benefits over classical methods for processing human data.

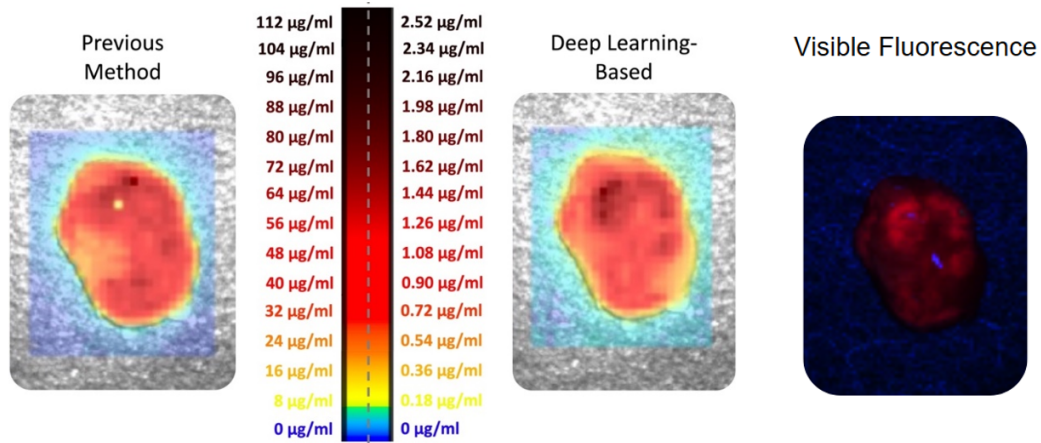

**Figure 7.** PpIX concentration computed across a brain tumor sample using ACU-SA (middle) and the classical method (left). The deep learning-based method shows a far more reasonable concentration range and better handles bright artifacts in the top center of the sample. The visible fluorescence shows very similar patterns to the unmixing results.

For ACU-Net, although we do not explicitly train to achieve a normalized fluorescence emission spectrum, we observe that the reconstructions do converge to a reasonable spectrum for samples of both phantom and pig brain homogenate given the same PpIX concentrations, as shown in Fig. 8.

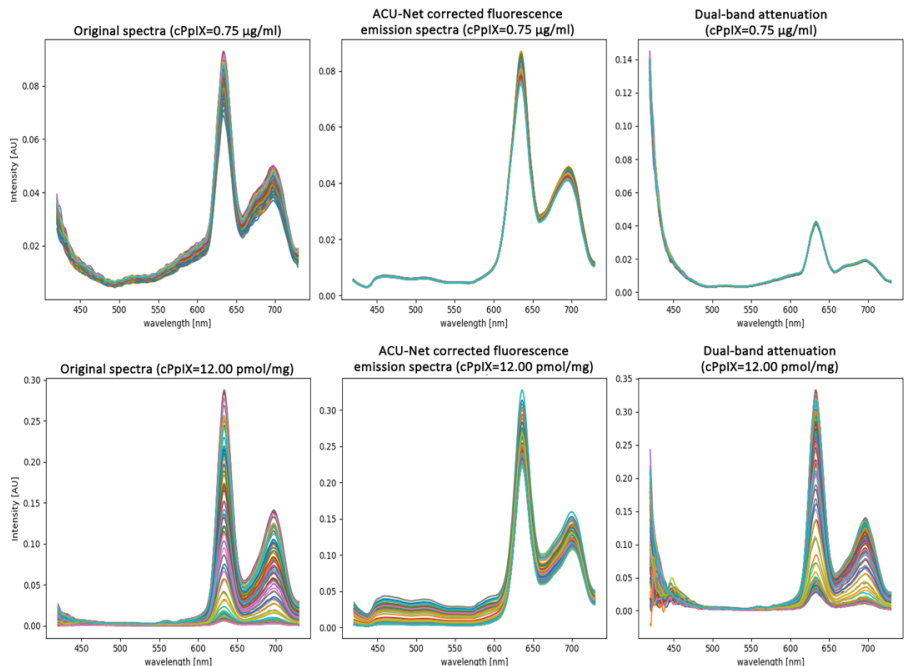

**Figure 8.** Corrected fluorescence emission spectra computed for both PPIX phantoms (top) and pig brain homogenate (bottom) using ACU-Net (middle) and the benchmark dual-band

attenuation method (right). The deep learning-based correction shows lower variance given the same concentration and better corrects for the blue light tail near the excitation wavelengths.

## Discussion

The results show that both supervised and semi-supervised learning outperform classical methods for correcting and unmixing hyperspectral brain tumor data. The performance of the unsupervised method is promising for the field, as it shows that improved performance may be achieved without labeled datasets. Instead, data such as our human measurements can be used without ground truth abundance values. In this way, such models could be trained with large volumes of data and may generalize well to new human measurements. Further work is required to continue improving the performance and to show quantitatively that it is effective on human data. This may involve chemical or histopathological assessment of samples co-registered to the HSI measurements, allowing for comparison of known absolute PpIX concentrations.

For better generalizability and interpretability, it is best to separate the normalization and unmixing steps or at least have an intermediate state, which is the normalized spectra. Then, for example, the normalization network could be trained on phantom data with concentration labels and then attached to an unmixing network, which was trained and unsupervised on human data. In this way, the whole model would generalize better since the unmixing cannot be trained on phantom data, which contains different endmember spectra than human brain, and the normalization is best trained with phantoms of constant, known concentration. This is achieved to a degree in this study but requires further investigation. Although ACU-Net and ACU-SA both achieved similarly high R value, we observed cases where the intermediate predicted normalized spectrum did not resemble a real measured spectrum. This indicates that the domain of the HU function the deep learning models learn is too large. Future work should find methods to constrain the shape of the predicted normalized spectrum more strongly to prevent the ACU-SA architecture from functioning as an end-to-end model and defeating the purpose of having a distinct normalization module. A related challenge is that the normalized spectrum is not known a priori. This is why both ACU-Net and ACU-SA rely on either an indirect or latent representation during training, which is not guaranteed to converge to true physical normalized spectrum. If phantoms are not sufficiently homogenous, the assumption that a common normalized spectrum exists is tenuous.

This study on the use of deep learning for analysis of hyperspectral images in fluorescence-guided neurosurgery invites several avenues of future research. These include integrating increasingly sophisticated models emerging from deep learning research, adding further constraints to enhance modeling accuracy, and enriching the available data sets to bolster the effectiveness of models. For example, to enable supervised learning on human data, mass spectrometry could be used to determine ground truth labels. It is also likely that spatial interactions between adjacent pixels in the hyperspectral images, which are currently not accounted for in our models, exist. Notably, relevant studies, including those using deep learning models for unmixing and otherwise analyzing hyperspectral images, have demonstrated improved performance when examining larger image regions instead of individual pixels. Therefore, adopting a model similar to ACU-SA but using a 2D CNN to account for spatial information can enhance correctional capabilities. This would likely improve the spatial smoothness of the abundance overlay plots and better handle localized artifacts such as bright reflections, as shown in Fig. 7.

Another promising direction for future research is the integration of product and quotient relations into deep learning models. Previous studies[18,64,65], have successfully utilized scaling factors that multiply or divide the measured fluorescence emission spectra for normalization. However, standard DNNs are better at capturing additive and non-linear relationships rather than direct multiplicative or divisive interactions. Incorporating multiplication or division operations directly or explicitly transforming them to log space into the model's architecture could enable a DNN to represent these simpler analytical relations more efficiently, thus reducing the likelihood of overfitting and potentially offering a more accurate model of reality. However, it is essential to exercise caution regarding non-differentiability when incorporating these operations, as they can pose challenges in the gradient-based optimization process typically used in training deep neural networks.

## Conclusion

This paper has introduced two deep learning architectures that outperform prior methods for attenuation correction and unmixing of hyperspectral images in fluorescence-guided brain tumor surgery. The models explicitly enforce adherence to physical models of the system and condition on prior knowledge of the present endmember spectra, thus retaining some of the reliability and explainability of classical methods. Furthermore, the second introduced architecture can be trained in a semi-supervised or unsupervised manner, which allows the use of unlabeled human data and encourages better generalizability. These models will hopefully enable more accurate classification of brain tumors and tumor margins for intraoperative guidance in future work.

## Contributors

Conception and Design: D.B., A.X., J.G., E.S.M.
Acquisition of data: S.Ka., E.S.M.
Statistical Analysis and Interpretation: D.B., A.X., J.G., B.L., E.S.M.,
Drafting the article. D.B., A.X., J.G.
Critically revising the article: all authors
Technical support: A.D.I.
Study supervision: E.S.M.

## Acknowledgements

We would like to thank Carl Zeiss Meditec AG (Oberkochen, Germany) for providing us with the OPMI pico system and the BLUE 400 filter.